\documentclass{aastex}
\usepackage{spr-astr-addons}
\usepackage{url}

\RequirePackage{color}

\begin{document}

\title{Thermal instability in  ionized plasma}
\shorttitle{Thermal instability in  ionized plasma}
\shortauthors{Shadmehri et al.}

\author{Mohsen Shadmehri\altaffilmark{1}}
\affil{Department of Mathematical Physics, National University of Ireland Maynooth, Maynooth, Co. Kildare, Ireland\\mshadmehri@thphys.nuim.ie }
\and \author{Mohsen Nejad-Asghar\altaffilmark{2}}
\and
\author{Alireza Khesali\altaffilmark{2}}
\affil{Department of Physics, Mazandaran University,
Babolsar,Iran}

\begin{abstract}
We study magnetothermal instability in the ionized plasmas including the effects of Ohmic, ambipolar and Hall diffusion. Magnetic field in the single fluid approximation does not allow transverse thermal condensations, however, non-ideal effects highly diminish the stabilizing role of the magnetic field in thermally unstable plasmas. Therefore, enhanced growth rate of thermal condensation modes in the presence of the diffusion mechanisms speed up the rate of structure formation.

\end{abstract}



\section{Introduction}
\label{sec:1}
 Role of the magnetic field in the dynamics of the gaseous astrophysical systems is generally studied within the framework of {\it ideal} MHD equations. In this simplified approach, which is good under certain conditions and circumstances, it is assumed that the coupling between the charged and neutral species of the system is perfect. But the perfect coupling assumption can be violated, in particular, when the density of the charged particles can be much lower than that of the neutral species. For example, one should note to this fact in dense molecular clouds \citep[e.g.,][]{ciolek2002}. Some authors have also criticized application of ideal MHD equation in modeling accretion discs, particular for the discs around young stellar objects \citep[e.g.,][]{wardle2004}.

The very existence of various species  with different masses and electrical charges and their collisions and the possible momentum transfer between the particles should be considered in any theory of structure formation in interstellar medium (ISM). However, interest towards modeling astrophysical plasmas within a multifluid approach has been raised over recent years \citep[e.g.,][]{falle03, turlough07, galli08, pandey2008, inutsuka2008}. In particular, simulating multifluid systems is a challenging area. For example, \cite{kim09}  incorporating ambipolar diffusion in the strong coupling approximation into a multidimensional magnetohydrodynamic (MHD) code based on the total variation diminishing scheme. More generalized Multifluid numerical schemes are also studied during recent years by some authors \citep[e.g.,][]{falle03, turlough07}.

On the theoretical side, efforts to understand physics of multifluid plasmas are in progress \citep[e.g.,][]{tassis, kunz09}. Recently,  \cite*{pandey2008} (hereafter PW) clarified the relationship between the fully ionized and weakly ionized limits by developing a unified single-fluid framework for the dynamics of the plasmas of arbitrary ionization. In another related study, general expressions for the resistivities, the diffusion timescales and the heating rates in a three-fluid medium are obtained by \cite*{galli08}.  They showed that the value of the Ohmic resistivity is increased in a collapsing cloud and the ambipolar diffusion occurs on a time scale comparable to the dynamical time scale.

Among various physical mechanisms responsible for the density inhomogeneities in the ISM, it has been realized for a long time that thermal instability can be an efficient processes \citep[e.g.,][]{field65,burkert00,heneb07,inutsuka2008,gazol09}. Thermal instability in a cooling medium can also be affected when dynamics of different charged species (e.g., dust particles) are included \citep{ibanez02}. It is shown that negatively charged particles stimulate the thermal instability in the sense that the conditions for the instability to hold are wider than similar conditions in a single-fluid description \citep{Kopp,shukla03}.

Thermal instability within ambipolar regime also studied by \cite*{asghar03} and \cite*{asghar07}, where in the former the frictional heating by the ion-neutral is included. In another similar study, \cite*{fabian06} revisited the problem of clump formation due to thermal instabilities in a weakly ionized plasma. However, dynamics of ions and their contribution to the net cooling function are not included in these studies of thermal instability in ambipolar regime. Therefore, \cite*{fukue07} extended the linear classical thermal instability to a case, in which dynamics of neutrals and ions and their interactions are considered. They showed that ion-neutral friction with the magnetic field affects the morphology and evolution of the interstellar matter. Thermal instability for a system obeying generalized Ohm's law has also been studied by \cite*{bora}. They found that the instability criterion involves the field strength, resistivity and electron inertia terms for transverse perturbations, but for the parallel to the ambient magnetic field the instability criterion is independent of all these non-ideal effects. However, they did not apply their analysis to the structure formation due to the thermal instability in the interstellar medium.

In this paper, our goal is to include non-ideal effects in a magnetized thermally unstable plasma. We follow a recent approach by PW, in which ions, electrons and neutrals are included. Then, non-ideal Ohmic, Hall and ambipolar terms appear in a generalized form of the induction equation. We study role of these terms in magnetothermal  instability in the linear regime. The basic assumptions and the equations are presented in the next section.  Linearized equations and the dispersion relation are derived in section \ref{sec:Linear}. Analyzing effects of non-ideal mechanisms on the thermally unstable modes are done in section \ref{sec:Analysis}. We conclude by a summary of the results and possible implications in the final section.

\section{Basic multifluid equations}
Basic assumptions and the equations of our model  are constructed based on the approach of PW. Here, an energy equation is introduced as well. The possible effects of charged dust particles on the thermal instability are neglected for simplicity \citep[e.g.,][]{shadmehri09}. We assume that the system is consisted of ions, electrons and neutrals. However, it is difficult to consider the full set of dynamical equations for a multifluid system with three different species. But one can reduce the set of  multifluid equations to a more manageable set of equations with some extra terms due to the non-ideal effects as has been done by some authors (e.g., PW). The basic equations of our analysis are discussed here, but one can refer to PW for a detailed discussion about the assumptions and the key approximations.

For each component of the system, the continuity equation is written as
\begin{equation}\label{con1}
    \frac{\partial \rho_{j}}{\partial t} + \nabla \cdot (\rho_{j} \textbf{v}_{j}) =
    0,
\end{equation}
where $\rho_{j}=m_{j} n_{j}$ is the mass density, $\textbf{v}_{j}$ is the
velocity, and $n_{j}$ and $m_{j}$ are the number density and particle
mass of ions, electrons and neutrals, i.e. $j={\rm i, e, n}$. The continuity
equation for the bulk fluid is obtained by summing up equation
(\ref{con1}) for each species. Therefore,
\begin{equation}\label{con2}
    \frac{\partial \rho}{\partial t} + \nabla \cdot (\rho \textbf{v}) = 0,
\end{equation}
where $\rho=\sum \rho_{j} \approx \rho_{\rm i}+\rho_{\rm n}$ and
$\textbf{v}=(\rho_{\rm i} \textbf{v}_{\rm i}+ \rho_{\rm n} \textbf{v}_{\rm n})/\rho$ are
the bulk fluid density and velocity, respectively.

We shall assume that the ions are singly charged and adopt charge
neutrality, i.e.  $n_{\rm i}=n_{\rm e}$. The momentum equations for the
electrons, ions and neutrals are
\begin{eqnarray}\label{momelec}
 \nonumber   \rho_{\rm e} (\frac{\partial \textbf{v}_{\rm e}}{\partial t}
 + \textbf{v}_{\rm e} \cdot \nabla \textbf{v}_{\rm e}) = -\nabla P_{\rm e} - n_{\rm e} e (\textbf{E} +
    \frac{\textbf{v}_{\rm e}}{c} \times \textbf{B}) \\
    - \rho_{\rm e} \sum_{j={\rm i,n}} \nu_{{\rm e}j} (\textbf{v}_{\rm e} -
    \textbf{v}_j),
\end{eqnarray}
\begin{eqnarray}\label{momion}
 \nonumber   \rho_{\rm i} (\frac{\partial \textbf{v}_{\rm i}}{\partial t}
 + \textbf{v}_{\rm i} \cdot \nabla \textbf{v}_{\rm i})  = -\nabla P_{\rm i} + n_{\rm i} e (\textbf{E} +
    \frac{\textbf{v}_{\rm i}}{c} \times \textbf{B}) \\
    - \rho_{\rm i} \sum_{j={\rm e,n}} \nu_{{\rm i}j} (\textbf{v}_{\rm i} - \textbf{v}_j),
\end{eqnarray}
\begin{equation}\label{momneut}
    \rho_{\rm n} (\frac{\partial \textbf{v}_{\rm n}}{\partial t}
    + \textbf{v}_{\rm n} \cdot \nabla \textbf{v}_{\rm n}) = -\nabla P_{\rm n} + \rho_{\rm n} \sum_{j={\rm e,i}} \nu_{{\rm n}j}
    (\textbf{v}_{j} - \textbf{v}_{\rm n}),
\end{equation}
respectively. The electron and ion momentum equations
(\ref{momelec})-(\ref{momion}) contain on the right hand side
pressure gradient, Lorentz force and collision momentum exchange
terms where $P_j$ is the pressure, $\textbf{E}$ and $\textbf{B}$
are the electric and magnetic field, $c$ is the speed of light,
and $\nu_{jk}$ is the collision frequency for $j$th component with $k$th component (i.e., $\rho_j \nu_{jk} = \rho_k \nu_{kj}$).

The bulk momentum equation can be derived by adding equations
(\ref{momelec})-(\ref{momneut}) to obtain
\begin{equation}\label{mombulk1}
  \rho ( \frac{\partial \textbf{v}}{\partial t} + \textbf{v} \cdot \nabla \textbf{v}) + \nabla \cdot (\frac{\rho_{\rm i} \rho_{\rm n}}{\rho} \textbf{v}_{\rm D}
  \textbf{v}_{\rm D})= - \nabla P +\frac{1}{c} \textbf{J} \times
  \textbf{B},
\end{equation}
where $P=P_{\rm e}+P_{\rm i}+P_{\rm n}$ is the total pressure,
$\textbf{v}_{\rm D}=\textbf{v}_{\rm i} - \textbf{v}_{\rm n}$ is the ion-neutral
drift velocity, and $\textbf{J}=n_{\rm e} e (\textbf{v}_{\rm i} -
\textbf{v}_{\rm e})$ is the current density. PW showed that the $\textbf{v}_{\rm D}
\textbf{v}_{\rm D}$ term in equation (\ref{mombulk1}) can be neglected
for dynamical frequencies satisfying

\begin{equation}\label{freq}
  \omega < \frac{\rho}{\sqrt{\rho_{\rm i} \rho_{\rm n}}} (\frac{D \beta_{\rm e}}{1+D \beta_{\rm e}})
  \nu_{\rm ni},
\end{equation}
where $D=\rho_{\rm n}/\rho$ is the neutral density fraction and
$\beta_{\rm e}$ is the ratio of the cyclotron frequency of the electron $\omega_{\rm e}$
to the sum of the electron-ion and electron-neutral collision
frequency $\nu_{\rm e}$ (PW), i.e. $\beta_{\rm e}=\omega_{\rm e}/\nu_{\rm e}$ or
\begin{equation}
\beta_{\rm e}=\frac{(eB/m_{\rm e}c)}{\nu_{\rm en}+\nu_{\rm ei}}.
\end{equation}
Then, we can recover the single-fluid momentum equation as
\begin{equation}\label{mombulk2}
  \rho (\frac{\partial \textbf{v}}{\partial t} + \textbf{v} \cdot \nabla \textbf{v}) = - \nabla P +\frac{1}{c} \textbf{J} \times
  \textbf{B}.
\end{equation}

To obtain an equation for the evolution of the magnetic field, we
need to drive an expression for the electric field $\textbf{E}$
in terms of the fluid properties to insert into Faraday's law. We present the main steps for obtaining the final form of the induction equation following detailed and extensive calculations of PW. First, we obtain a relation for
 ${\bf v}_{\rm D}$ by rewriting the ion and neutral equations of motion (\ref{momelec}) and (\ref{momion}) as
\begin{displaymath}
(\rho_{\rm i}\nu_{{\rm in}}+\rho_{\rm e} \nu_{\rm en}) {\textbf{v}_{\rm D}} = - \rho_{\rm i} (\frac{\partial {\textbf{v}_{\rm i}}}{\partial t} + {\bf v}_{\rm i}. \nabla {\bf v}_{\rm i})
\end{displaymath}
\begin{equation}\label{eq:reff1}
-\nabla (P_{\rm e} + P_{\rm i}) + \frac{{\bf J}\times {\bf B}}{c} + \frac{m_{\rm e}\nu_{\rm en}}{e} {\bf J},
\end{equation}
and
\begin{displaymath}
(\rho_{\rm i}\nu_{{\rm in}}+\rho_{\rm e} \nu_{\rm en}) {\textbf{v}_{\rm D}} = - \rho_{\rm n} (\frac{\partial {\textbf{v}_{\rm n}}}{\partial t} + {\bf v}_{\rm n}. \nabla {\bf v}_{\rm n})
\end{displaymath}
\begin{equation}\label{eq:reff2}
+\nabla P_{\rm n} +  \frac{m_{\rm e}\nu_{\rm en}}{e} {\bf J}.
\end{equation}
Multiplying equation (\ref{eq:reff1}) by $\rho_{\rm n}$ and equation (\ref{eq:reff2}) by $\rho_{\rm i}$ and then adding and noting that
 $\rho_{\rm e} \nu_{\rm en} \ll \rho_{\rm i} \nu_{\rm in}$, we obtain
\begin{equation}\label{eq:VD}
{\bf v}_{\rm D}=D \frac{{\bf J}\times {\bf B}}{c \rho_{\rm i} \nu_{\rm in}} + \frac{\nabla P_{\rm n}}{\rho_{\rm i}\nu_{\rm in}} - D \frac{\nabla P}{\rho_{\rm i} \nu_{\rm in}} + (\frac{\beta_{\rm i}}{\beta_{\rm e}}) \frac{\bf J}{e n_{\rm e}}.
\end{equation}
Obviously, the above equation is not valid when $\nu_{\rm in}$ tends to zero unless we relax the assumption that we used, i.e. $\rho_{\rm e} \nu_{\rm en} \ll \rho_{\rm i} \nu_{\rm in}$. In the weakly ionized limit, equation (\ref{eq:VD}) reduces to the strong coupling approximation, i.e. ${\bf v}_{\rm D} \approx  ({\bf J}\times {\bf B})/(c \rho_{\rm i} \nu_{\rm in})$.

Now, we need a relation for the electric field. The electron momentum equation (\ref{momelec}), in the zero
electron inertia limit, yields an expression for the electric
field in the rest frame of the ions
\begin{equation}\label{electric}
  \textbf{E} = - \frac{\textbf{v}_{\rm i}}{c} \times \textbf{B} - \frac{\nabla P_{\rm e}}{e
  n_{\rm e}}+ \frac{\textbf{J}}{\sigma} + \frac{\textbf{J} \times
  \textbf{B}}{cen_e}- \frac{m_{\rm e}\nu_{\rm en}}{\rm e} \textbf{v}_{\rm D},
\end{equation}
where $\sigma= e^2 n_{\rm e} / m_{\rm e} (\nu_{\rm en} + \nu_{\rm ei})$ is the Ohmic
conductivity and $\textbf{J}$ is given by Ampere's law,
$\textbf{J}=(c/4\pi) \nabla \times \textbf{B}$. But we note ${\bf v}_{\rm i} = {\bf v} + D {\bf v}_{\rm D}$. Substituting equation (\ref{eq:VD}) into equation (\ref{electric}), the induction equation $\partial {\bf B} / \partial  t = -c \nabla\times {\bf E}$ becomes
(PW)
\begin{eqnarray}\label{magbulk}
 \nonumber  \frac{\partial \textbf{B}}{\partial t} = \nabla \times [ (\textbf{v} \times \textbf{B})
   - \frac{4\pi\eta}{c} \textbf{J} - \frac{4\pi\eta_{\rm H}}{c} \textbf{J} \times \hat{\textbf{B}}
  \\ +\frac{4\pi\eta_{\rm A}}{c} (\textbf{J} \times \hat{\textbf{B}}) \times \hat{\textbf{B}} ],
\end{eqnarray}
where $\hat{\textbf{B}}=\textbf{B}/B$ and the Ohmic, Hall and ambipolar coefficients are $ \eta=c^2/4\pi\sigma$,
$\eta_{\rm H}=cB/4\pi e n_{\rm e}$ and $\eta_{\rm A}=D^2 B^{2}/4\pi \rho_{\rm i}
   \nu_{\rm in}$, respectively.

Now, if we write the energy equation  for ions, electrons and neutrals and
add up the equations, and neglecting the external cooling
and heating of electrons, we obtain the bulk energy
equation that is given by its usual form as follows
\begin{displaymath}
    \frac{1}{\gamma -1} (\frac{\partial P}{\partial t}+ {\bf v}\cdot\nabla P)
    +\frac{\gamma}{\gamma -1} P \nabla\cdot{\bf v}
\end{displaymath}
\begin{equation}\label{energy2}
+\rho\Omega - \nabla . [K_{\parallel} \nabla_{\parallel} T + K_{\perp} \nabla_{\perp} T]= 0,
\end{equation}
where $\Omega$ is the net cooling function. Also, the coefficient of thermal conductivity $K$ has the values $K_{\parallel}$ and $K_{\perp}$ in directions parallel and perpendicular to the magnetic field ${\bf B}$.

There are some points regarding the energy equation (\ref{energy2}) and the net cooling function, in particular. In analysing thermal instability, the net cooling function, $\Omega$, has a vital role \citep[e.g.,][]{field65}. In addition to the usual heating-cooling terms, three extra heating terms appear in the net cooling function because of the non-ideal effects, notably Ohmic, Hall and Ambipolar. While  the usual heating-cooling terms are actually local functions and depend only on the physical quantities of the system, non-ideal heating terms are determined mainly by  the {\it spatial} variations of the magnetic field. In other words, all non-ideal heating terms vanish for a system with uniform magnetic field. In such circumstances, equilibrium states of the system are solely determined by the local heating-cooling terms. It means that the classical criteria of thermal instability is still applicable when the magnetic field is uniform. But growth rate of the thermally unstable  modes may significantly modify due to the non-ideal effects on the distribution of magnetic field through our modified induction equation.  On the other hand, linear analysis of thermal instability is restricted to keeping terms which are of the first order and since  non-ideal heating terms are of the second order, they do not appear in the linearized set of the equations, irrespective of the profile of the magnetic field. Thus, possible effects of non-ideal effects on the growth rate of thermally unstable modes appears through modified induction equation in our simplified analysis which is restricted to the linear regime with  uniform initial conditions.

Therefore, equations (\ref{con2}), (\ref{mombulk2}), (\ref{magbulk}) and (\ref{energy2}) along with the ideal gas equation of state, $p=(R/\mu ) \rho T$, are the main equations of our model in order to study thermal instability in a multifluid system. The equations look like similar to the ideal MHD equations, except for the induction equation where some extra terms emerge due to the non-ideal effects such as ion-electron or ion-neutral drift velocities.

\section{Linear perturbations}
\label{sec:Linear}

In the local homogeneous equilibrium state, we have $\rho=\rho_0$,
$P=P_0$, $T=T_0$, $\textbf{B}= \textbf{B}_0$, $\textbf{v}=0$, and
$\Omega(\rho_0,T_0)=0$. We assume perturbations of the form
\begin{equation}\label{perturb}
\chi (\textbf{r},t) = \chi_1 \exp
 (\omega t + i \textbf{k} \cdot \textbf{r}),
\end{equation}
where $\chi_1$ is the amplitude of the perturbations,
$\omega$ is the growth rate of the perturbations and ${\bf k}$ is the wavenumber of the perturbations. We are interested in modes where $\omega$ is real and positive, i.e.
condensation modes.

Then the linearized  equations are
\begin{equation}\label{masslin}
\omega\rho_1 + i\rho_0 \textbf{k} \cdot \textbf{v}_1=0,
\end{equation}
\begin{equation}\label{momenlin}
\omega \rho_0 \textbf{v}_1+ i \textbf{k}  P_1+ i
(\textbf{B}_0\cdot\textbf{B}_1) \frac{\textbf{k}}{4\pi}-
i(\textbf{k}\cdot\textbf{B}_0)\frac{\textbf{B}_1} {4\pi}=0,
\end{equation}
\begin{displaymath}
\frac{\omega}{\gamma-1}P_1-\frac{h\gamma P_0}{(\gamma-1)\rho_0}\rho_1+
\rho_0 \Omega_\rho \rho_1 +\rho_0 \Omega_T T_1
\end{displaymath}
\begin{equation}\label{energlin}
+(K_{\parallel} k_{\parallel}^{2} + K_{\perp} k_{\perp}^{2}) T_{1}=0,
\end{equation}
\begin{eqnarray}\label{maglin}
\nonumber \omega \textbf{B}_1 +i \textbf{B}_0 (\textbf{k} \cdot
\textbf{v}_1) - i (\textbf{k} \cdot\textbf{B}_0) \textbf{v}_1 -
\eta \textbf{k} \times (\textbf{k} \times \textbf{B}_1) \\
\nonumber + \eta_A \textbf{k} \times \{ [(\textbf{k} \times
\textbf{B}_1) \times \textbf{B}_0 ] \times \textbf{B}_0 \} \\ -
\eta_H (\textbf{k} \cdot \textbf{B}_0) (\textbf{k} \times
\textbf{B}_1)=0,
\end{eqnarray}
\begin{equation}\label{idealin}
  \frac{P_1}{P_0} - \frac{\rho_1}{\rho_0} - \frac{T_1}{T_0}=0,
\end{equation}
where the derivative $\Omega_{\rho}=(\partial\Omega/\partial\rho)_{T}$ and $\Omega_{T}=(\partial\Omega/\partial T)_{\rho}$ are evaluated for the equilibrium state.

We introduce the coordinate system $\textbf{\textit{e}}_x$,
$\textbf{\textit{e}}_y$, and $\textbf{\textit{e}}_z$ specified by
\begin{equation}\label{coord}
\textbf{\textit{e}}_z=\frac{\textbf{B}_0}{B_0}\quad,\quad\textbf{\textit{e}}_y=\frac{\textbf{B}_0\times\textbf{k}}
{|\textbf{B}_0\times\textbf{k}|}\quad,\quad\textbf{\textit{e}}_x=\textbf{\textit{e}}_y\times\textbf{\textit{e}}_z.
\end{equation}

Also, we introduce the following wavenumbers
\begin{displaymath}
k_{\rho}=\mu (\gamma -1) \rho_{0}\Omega_{\rho} (Rc_{\rm s} T_{0})^{-1},
\end{displaymath}
\begin{displaymath}
k_{T} = \mu (\gamma -1 ) \Omega_{T} (Rc_{\rm s})^{-1},
\end{displaymath}
\begin{displaymath}
k_{K_{\parallel}} = [\mu (\gamma -1) K_{\parallel}]^{-1} (Rc_{\rm s}\rho_{0}),
\end{displaymath}
\begin{equation}
k_{K_{\perp}} = [\mu (\gamma -1) K_{\perp}]^{-1} (Rc_{\rm s}\rho_{0}).
\end{equation}
Now, we can write the dispersion equation using the following non-dimensional quantities,
\begin{equation}
\sigma_{\rho} = \frac{k_{\rho}}{k}, \sigma_{T}=\frac{k_{T}}{k}, \sigma_{K_{\parallel}}=\frac{k}{k_{K_{\parallel}}}, \sigma_{K_{\perp}}=\frac{k}{k_{K_{\perp}}}.
\end{equation}
Our study differs from the classical thermal instability analysis \citep{field65} in introducing a generalized form for the induction equation because of the non-ideal effects. For numerical purposes, it is more convenient, to re-write Ohmic, ambipolar and Hall coefficients in terms of some
 non-dimensional parameters. In doing so, we have
\begin{equation}
\eta = \frac{c_{s}}{k_{\rho}} {\cal O},
\end{equation}
\begin{equation}
\eta_{\rm A}=\frac{c_{s}}{k_{\rho}} {\cal A},
\end{equation}
\begin{equation}
\eta_{\rm H}=\frac{c_{s}}{k_{\rho}} {\cal H},
\end{equation}
where the non-dimensional parameters ${\cal O}$, ${\cal A}$ and ${\cal H}$ are defined as
\begin{equation}\label{eq:O}
{\cal O}= \alpha \left [ \frac{k_{\rho} c_{s}}{\beta_{e} (m_{\rm e}/m_{\rm i}^{*}) \omega_{ce}} \right ],
\end{equation}
\begin{equation}\label{eq:A}
{\cal A}=\alpha D (\frac{k_{\rho} c_{s}}{\nu_{\rm ni}}),
\end{equation}
\begin{equation}
{\cal H}=\alpha (\frac{k_{\rho}c_{s}}{\omega_{\rm H}}).
\end{equation}
where $\alpha=(v_{\rm A}/c_{\rm s})^{2}$ and $v_{\rm A}$ is the Alfven velocity. Here, the effective ion mass is $m_{\rm i}^{\ast}=\rho / n_{\rm e}$. Then, the Hall frequency is defined as $\omega_{\rm H}=eB/m_{i}^{\ast}c$. Also, the cyclotron frequency of electrons is written as $\omega_{\rm ce}=eB/m_{\rm e}c$.

Therefore, the characteristic equation becomes
\begin{eqnarray}\label{eq:disper}
\nonumber Y^{7}+P_{6}Y^{6}+P_{5}Y^{5}+P_{4}Y^{4}+P_{3}Y^{3}+P_{2}Y^{2}
\\ +P_{1}Y+P_{0}=0,
\end{eqnarray}
where $Y=\omega/k_{\rho}c_{\rm s}$ and  the coefficients  are
\begin{equation}
P_{0}=\frac{\alpha^2 \xi^{2}}{\gamma} (\sigma_{T}+\sigma_{K}-\sigma_{\rho}),
\end{equation}
\begin{eqnarray}
P_{1}=\frac{\alpha\xi}{\gamma} (\sigma_{T}+\sigma_{K}-\sigma_{\rho}) [2{\cal O}+(\xi +1){\cal A}]+\alpha^{2}\xi^{2},
\end{eqnarray}
\begin{eqnarray}
\nonumber P_{2}=\frac{\xi}{\gamma} (\sigma_{T}+\sigma_{K}-\sigma_{\rho}){\cal A}^{2} + [ (\frac{1+\xi}{\gamma})(\sigma_{T}+\sigma_{K}
\\ \nonumber -\sigma_{\rho}){\cal O} + \alpha\xi (\xi +1) ] {\cal A} + \frac{1}{\gamma} (\sigma_{T}+\sigma_{K}-\sigma_{\rho}) {\cal O}^{2}
\\ \nonumber 2\alpha\xi {\cal O} + \frac{\xi}{\gamma} (\sigma_{T}+\sigma_{K}-\sigma_{\rho}) {\cal H}^{2} + \alpha \xi (\alpha + \frac{2}{\gamma})
\\ \times (\sigma_{T}+\sigma_{K})-\frac{2\alpha\xi}{\gamma}\sigma_{\rho}
\end{eqnarray}
\begin{eqnarray}
\nonumber P_{3}=\xi {\cal A}^{2} + [(1+\xi){\cal O} + (2\alpha\xi + \frac{1+\xi}{\gamma})(\sigma_{T}+\sigma_{K})
\\ \nonumber -(\frac{1+\xi }{\gamma})\sigma_{\rho}]{\cal A} + {\cal O}^{2} + [(\alpha\xi+\alpha+\frac{2}{\gamma})(\sigma_{T}+\sigma_{K})
\\ -\frac{2\sigma_{\rho}}{\gamma}]{\cal O} + \xi {\cal H}^{2} + \alpha\xi (\alpha +2)
\end{eqnarray}
\begin{eqnarray}
\nonumber P_{4}= \xi (\sigma_{T}+\sigma_{K}) {\cal A}^{2} + [ (1+\xi)(\sigma_{T}+\sigma_{K}){\cal{O}}+2\alpha\xi
\\ \nonumber + \xi + 1 ] {\cal A} + (\sigma_{T}+\sigma_{K}) {\cal O}^{2} + (2+\alpha + \alpha\xi) {\cal O}
\\ +\xi (\sigma_{T}+\sigma_{K}) {\cal H}^{2} + (\alpha + \alpha \xi + \frac{1}{\gamma}) (\sigma_{T}+\sigma_{K}) - \frac{\sigma_{\rho}}{\gamma}
\end{eqnarray}
\begin{eqnarray}
\nonumber P_{5}=\xi {\cal A}^{2} + (1+\xi )({\cal O} +\sigma_{T}+\sigma_{K} ) {\cal A} + {\cal O}^{2}
\\ +2(\sigma_{T}+\sigma_{K}){\cal O} + \xi {\cal H}^{2}+\alpha + \alpha\xi + 1
\end{eqnarray}
\begin{equation}
P_{6}=(1+\xi){\cal A}+2{\cal O}+\sigma_{T}+\sigma_{K}.
\end{equation}
and $\xi = \cos^{2}\theta$ and $\theta$ is the angle between ${\bf B}_{0}$ and ${\bf k}$. Also, we have $\sigma_{K}=\sigma_{K_{\parallel}} \xi + \sigma_{K_{\perp}} (1-\xi)$. Our analysis is based on dispersion equation (\ref{eq:disper}) and its roots are analyzed in the next section.

\section{Analysis}
\label{sec:Analysis}

If we neglect the non-ideal effects and set ${\cal O}={\cal A}={\cal H}=0$, the algebraic equation (\ref{eq:disper}) reduces to the standard magneothermal instability characteristic equation \citep{field65}. It is hard to do Hurwitz analysis in order to study all possible roots of the dispersion equation qualitatively due to its complicated coefficients. But equation (\ref{eq:disper}) is of odd degree in $Y$ and must therefore admit at least one positive real root for non-perpendicular perturbation if the last term $P_0$ is negative. Obviously, a positive root implies monotonic instability, i.e. condensation mode. Therefore, as long as different species doe not contribute to the net cooling function the condition of instability is the same as the ideal case, i.e. $P_{0} <0$ or $(\sigma_{T}+\sigma_{K}-\sigma_{\rho})<0$.

 \begin{figure*}
\includegraphics{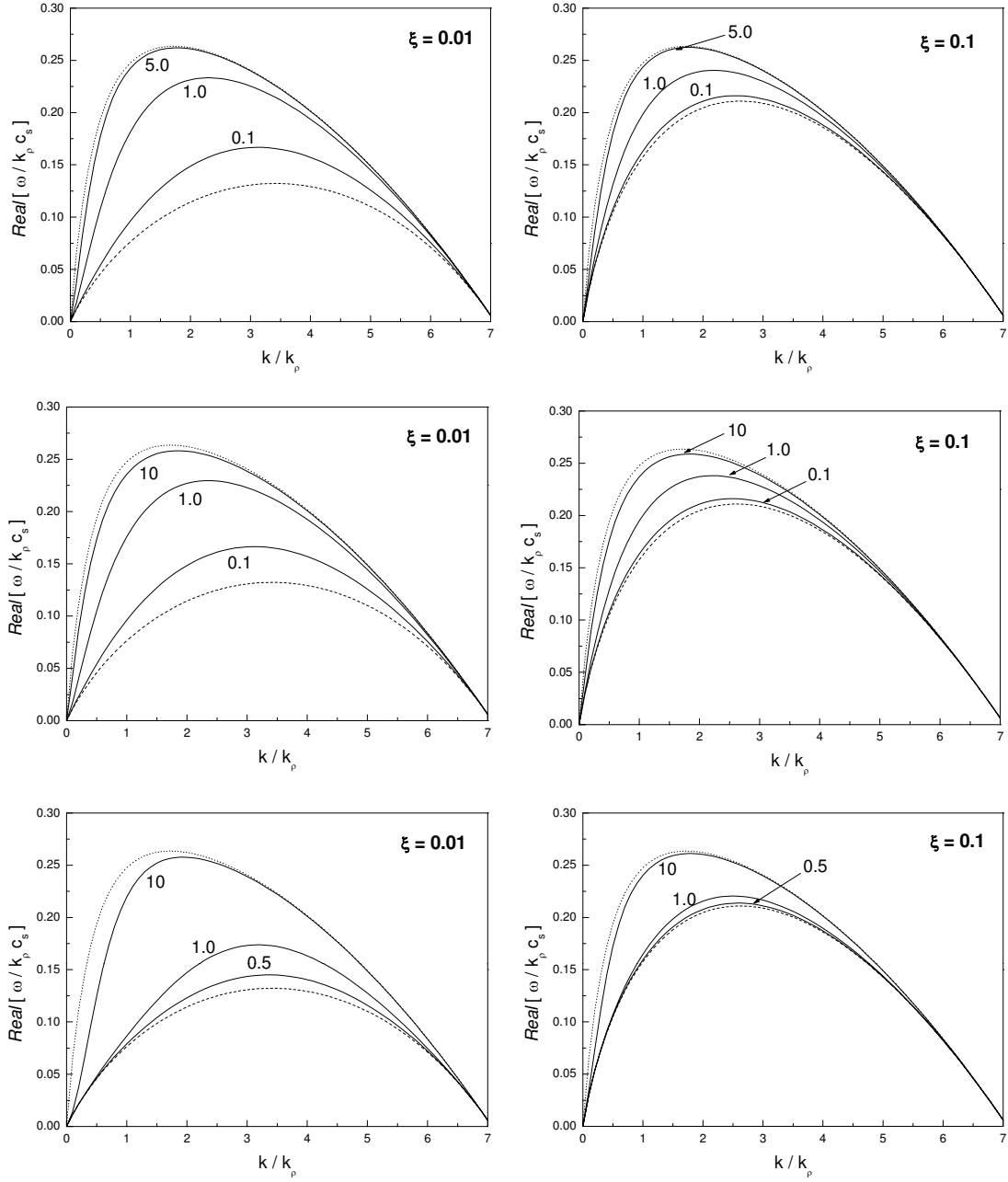}
\caption{Growth rates of magneto-thermal condensation mode versus wavenumber of the perturbations for three cases: Ohmic regime (top), Ambipolar regime (middle) and Hall regime (bottom). Our fiducial parameter are $\alpha=1$, $\gamma=5/3$, $\sigma_{\rm T}/\sigma_{\rho} = 1/2$ and $\sigma_{\rho} \sigma_{K} = 0.01$. Dotted and dashed curves are representing growth rates in the non-magnetized and ideal magnetized  cases. Each curve in the top, middle and bottom plots is labeled by fiducial values of ${\cal O}$, ${\cal A}$ and ${\cal H}$, respectively.}
\label{fig:1}
\end{figure*}

Also, we found that equation (\ref{eq:disper}) can not be factorized to a product of lower order polynomials unless the effect of Hall diffusion is neglected. If we set ${\cal H}=0$, the equation becomes a fifth order polynomial multiply by  a second order polynomial, i.e. $Y^{2}+({\cal O}+{\cal A}\xi )Y +\alpha \xi$. Then, the fifth order polynomial describes unstable modes and two stable waves modes are described by setting this second order polynomial equal to zero.

Now, we find real and positive root of equation (\ref{eq:disper}) numerically. We take the parameters $\gamma =5/3$, $\alpha=1$, $\sigma_{\rm T}/\sigma_{\rho} = 1/2$ and $\sigma_{\rho} \sigma_{K} = 0.01$ for comparison to Field (1965). Figure \ref{fig:1} shows growth rate versus the wavenumber of the perturbations for Ohmic, ambipolar and Hall regimes, separately. For easier comparison, growth rates in the non-magnetized and ideal magnetized  cases are shown by dotted and dashed curves, respectively. In the top plots of Figure  \ref{fig:1}, we set ${\cal A}={\cal H}=0$ and the non-dimensional Ohmic coefficient ${\cal O}$ varies from $0.1$ to $5$. Each curve is labeled by its corresponding ${\cal O}$. Growth rate of the condensation mode increases with the coefficient ${\cal O}$ and the effect is stronger for the transverse perturbations. Obviously, larger growth rate implies a more magnetothermally unstable system. So, Ohmic diffusion has a destabilizing effect on the condensation mode. As Ohmic coefficient becomes large, the profile of the growth rate tends to the non-magnetized case. In other words, the effect of magnetic field on the condensation mode diminishes in the presence of an efficient Ohmic diffusion.

Middle plots of Figure \ref{fig:1} show growth rate in ambipolar regime, i.e. ${\cal O}={\cal H}=0$ but ${\cal A}\neq 0$. Here, non-dimensional ambipolar diffusion coefficient ${\cal A}$ varies from $0.1$ to $10$. As in Ohmic regime, a destabilizing effect is seen due to ambipolar  dissipation. Bottom plots of Figure \ref{fig:1} show growth rate of the condensation mode when the system is in the Hall regime, i.e. ${\cal O}={\cal A}=0$ and ${\cal H}\neq 0$. Destabilizing effect of Hall diffusion on the thermal instability is amplified with the coefficient ${\cal H}$. The wavelength of perturbation that maximizes the growth rate profile shifts to longer wavelengths as non-ideal effects become  stronger. Then, larger clouds are the most likely structures to be formed due to magnetothermal instability in comparison to the ideal case when non-ideal effects are taken into account.

Our results are explained using a few simple points. Different species not only may contribute to the net cooling of the system but their dynamical roles will appear either by direct interactions and momentum exchanges or through magnetic field forces for the charged particles.  However, the bulk density and velocity are mainly determined by the most massive particle, i.e. neutrals. Because electrons are very light and their dynamical roles can always be neglected. Moreover, in the weakly ionized case, the mass fraction of ions are small in comparison to the total density of the thermal gas.  In the ideal  MHD description, all the charged particles are well coupled to the magnetic field lines and the neutrals are moving along with the charged species. But in reality, magnetic coupling of the charged particles may vary depending on the mass of the particles and the magnetic strength. Obviously, electrons are more coupled to the magnetic field lines in comparison to the other charged particles. Thus, the strength of the magnetic force may change. In our model, response of the system to the perturbations is mainly determined by the distribution of the magnetic field lines. So, non-ideal effects lead to a reduction to the magnetic pressure and this makes the system more unstable.

Considering the above point, growth rates in Figure \ref{fig:1} are physically understandable. Growth rates corresponding to the non-magnetized and ideal magnetized cases are also shown in all panels of this Figure. We can consider two thermal and magnetized systems: one including non-ideal magnetic effects and the other one in the ideal magnetized case. Let's consider transverse perturbations, first. If we compare ideally magnetized  case with non-magnetized situation, we see that magnetic field stabilizes the system, i.e. growth rates decreases in the ideally magnetized case. It can be simply explained by the magnetic pressure that provides additional support against perturbations. Thus, in the ideally magnetized case, there are actually two types of pressures: thermal and magnetic. So, in order to have thermal instability, perturbations should overcome total pressure which consists of the thermal and the magnetic pressures. Now, we consider non-ideal effects such as resistivity or ambipolar. Obviously, these mechanisms are dissipative. It means that magnetic flux and eventually magnetic pressure is reduced due to the non-ideal effects. Thus, there is less total pressure in non-ideal case in comparison to the ideal case. This implies that growth rates are increased with non-ideal parameters and the curves are shifting upwards, i.e. from ideal case to the non-magnetized case.

Equation  (\ref{eq:A}) shows that the ambipolar coefficient is directly proportional to the ratio $D = \rho_{\rm n} / \rho$. So, small values of $\cal A$ correspond to when $D$ tends to zero. On the other hand, one can simply show that $\cal O$ is directly proportional to $D$ according to equation (\ref{eq:O}).  In a highly ionized case, i.e. $D$ tends to zero, unlike Ohmic and ambipolar mechanisms, Hall diffusion does not disappear in the induction equation. So, the only effect of $D$ is seen directly via parameters $\cal O$ and $\cal A$. Figure \ref{fig:1} shows an extensive study of the parameter space, and we can see simply that as $A$ or $O$ tends to zero the growth rates tend to the classical result of the thermal instability.

We can now apply our linear multifluid  magnetothermal instability to structure formation in ISM.  As an example, we consider structure formation in typical HI clouds \citep{wolfire}. In our calculation, we adopt $m_{\rm i}=30m_{\rm p}$ and $m_{\rm n}=2.33m_{\rm p}$ for the ion and mean neutral mass, respectively. Here, $m_{\rm p}=1.67\times 10^{-24}$ g is the proton mass.   We assume that the temperature of HI cloud is $100$ K. The number density of neutral component and the magnetic field are estimated to be $71.9$ cm$^{-3}$ and $10^{-6}$ G. Also, we have $\gamma=5/3$. Under these conditions, CII cooling is a dominant mechanism  according to \citet{wolfire} and the cooling rate depends on the ionization degree and the fraction of CI in CII (i.e., $f_{\rm CII}$). For example, for ionization degree $10^{-6}$ and $f_{\rm CII}=0.01$, the cooling rate becomes $k_{\rho} c_{\rm s}=9.28\times 10^{-14}$ s  \citep[e.g.,][]{fukue07}. Assuming that all ions are singly ionized, our diffusive non-dimensional parameters becomes ${\cal O}/\alpha = 3.4\times 10^{-10}$, ${\cal A}/\alpha = 9.4$ and ${\cal H}/\alpha = 0.35$. Thus, Ohmic dissipation is not operating in HI regions. But ambipolar and Hall diffusions are significant processes to be considered for HI structure formation due to magnetothermal instability.

\section{Summary and Conclusions}
We studied magnetothermal instability with the effect of non-ideal Ohmic, ambipolar and Hall diffusion. Our linear analysis shows that the criteria of instability does not change comparing to the ideal case as long as charged species do not contribute to the net cooling function. Also, the system becomes more unstable in the presence of non-ideal effects and it is more probable to have larger clouds comparing to the ideal case.    Although the vital role of magnetic field dissipation in very dense interstellar clouds is a key process in standard theories of star formation \citep[e.g.][]{nakano86}, our results show that such non-ideal mechanisms may operate in thermally unstable systems such as HI regions or warm ISM.

\cite*{inutsuka2008} studied formation of structures  in a weakly ionized and magnetized interstellar medium using two-fluid magnetohydrodynamic simulations. When orientation of magnetic field is perpendicular to the flow, the rate of formation of clouds slows down significantly according to \cite*{inutsuka2008}. Actually, linear ideal magnetothermal instability shows that magnetic field prevents structure formation transverse to the field lines \citep{field65}. However, our results show that the stabilizing effect of magnetic field drastically diminishes because of the dissipative processes like Ohmic or ambipolar. Therefore, we think, the opportunity of fast molecular cloud formation directly from the warm neutral medium would highly increase due to the non-ideal effects, at least in the linear regime.

One should note that enhanced growth rate due to the non-ideal terms is independent of the true mechanisms of the operating processes. But we think this independency is valid in the linear regime and the nonlinear evolution of the system will depend on the type of dominant diffusion process. In particular, Hall diffusion significantly differs from Ohmic and ambipolar diffusions regarding to the energy considerations. In fact, it is known that Ohmic and ambipolar diffusion are dissipative processes and reduce magnetic energy of the system. But Hall diffusion does not contribute to the dissipation of the magnetic energy and so, it is not a dissipative mechanism. Its main role goes back to redistributing the current within the system and it may lead to enhanced dissipation of the magnetic energy because of Ohmic or ambipolar diffusions. Numerical simulations show that the dissipation rate of the MHD turbulence is strongly affected by the strength of ambipolar diffusion \citep{kim09}. But it is an open question to explore possible role of Hall diffusion in nonlinear evolution of thermally unstable system \citep[see also,][]{wardle2004}.

\section*{Acknowledgments}
We are grateful to the anonymous referee whose comments helped to improve the quality of this paper.

\bibliographystyle{spr-mp-nameyear-cnd}
\bibliography{biblio-u1}

\end{document}